\documentclass[pre,twocolumn,english,showpacs,amsmath,amssymb,superscriptaddress,longbibliography]{revtex4-2}
\usepackage{color}
\usepackage{mathptmx}

\usepackage[T1]{fontenc}
\usepackage{float}
\usepackage{graphicx}
\usepackage[normalem]{ulem}
\usepackage{amsmath}
\usepackage{amssymb}
\usepackage{soul}
\usepackage{bm}
\usepackage{boldline,multirow}
\usepackage{stackrel}
\usepackage{comment}

\usepackage{bibunits}



\usepackage{babel}
\usepackage{bm}
\usepackage{xr}

\begin{document}



\renewcommand{\figurename}{\textbf{Fig.}}
\renewcommand{\thefigure}{\arabic{figure}}
\renewcommand{\thetable}{\arabic{table}}

\title{The structural order of protein hydration water}

\author{Rui Shi}
\email{ruishi@zju.edu.cn}
\affiliation{Zhejiang Province Key Laboratory of Quantum Technology and Device, School of Physics, \\
Zhejiang University, Zheda Road 38, Hangzhou 310027, China.}
\email{ruishi@zju.edu.cn}

\date{\today}
\vspace{-7mm}
\begin{abstract}
The ability of water to dissolve biomolecules is crucial for our life. It has been shown that protein has a profound effect on the behavior of water in its hydration shell, which in turn affects the structure and function of the protein. However, there is still no consensus on whether protein promotes or destroys the structural order of water in its hydration shell until today, because of the lack of proper structural descriptor incorporating hydrogen-bond (H-bond) information for water at the protein/water interface. Here we performed all-atom molecular dynamics simulations of lysozyme protein in water and analyzed the H-bond structure of protein hydration water by using a newly developed structural descriptor. We find that the protein promotes local structural ordering of the hydration water while has a negligible effect on the strength of individual H-bond. These findings are fundamental to the structure and function of biomolecules and provide new insights into the hydration of protein in water.

\end{abstract}

\maketitle

\section{Introduction}

Protein maintains its structure and function upon solvation in water. There is increasing evidence supporting that water not only acts as a solvent but also actively participates in many biological processes~\cite{ball2008water,ball2017water}. For examples, it has been shown that the protein hydration water has a significant impact on protein dynamics~\cite{heyden2013spatial,schiro2015translational,Mukherjee2019}, protein-ligand binding~\cite{beuming2012thermodynamic,spitaleri2020tuning}, protein stability~\cite{bianco2015contribution,kim2016computational,bianco2017role,kozuch2019low} and the catalytic efficiency of enzyme ~\cite{lather2020improving}. Therefore, the solvation of protein is key towards understanding the biological function of proteins.

Water being able to form connected hydrogen-bond (H-bond) network with locally favored tetrahedral symmetry is the most unique and anomalous solvent in nature~\cite{eisenberg2005structure,angell1983supercooled,debenedetti2003supercooled,gallo2016water}. It has been shown that the local tetrahedral ordering is responsible for both the thermodynamic and dynamic anomalous behaviors of water~\cite{shi2018origin,shi2018common,shi2020direct,shi2020anomalies}. The presence of solute inevitably perturbs the tetrahedral structure of water. In 1959 Kauzmann proposed that the water structural ordering around hydrophobic solutes is the origin of the hydrophobic interaction which serves as the key driving force of the protein folding and aggregating in aqueous solutions~\cite{kauzmann1959some}. It's now well accepted that the hydrophobic interaction is entropic in its origin, but how to characterize the underlying water structure in the vicinity of proteins has remained a major challenge so far. 

The structure of protein hydration water has been intensively studied by using various structural descriptors. However, either experiments or simulations report contradictory effects of protein on the hydration water structure. For example, Shen et al. reported that hydration water of most amino acids has higher tetrahedral order than bulk water by using Raman multivariate curve resolution spectroscopy~\cite{shen2021quantitative}. Enhanced H-bonding structure of protein hydration water has also been found by Fourier transform infrared spectroscopy~\cite{panuszko2012characteristics} and femtosecond surface sum frequency generation spectroscopy~\cite{meister2014observation}. These results are supported by molecular dynamics (MD) simulations that reported a significantly structured hydration water layer around a lysozyme protein~\cite{accordino2012temperature,camisasca2018structure}.

In contrast, neutron Brillouin measurements combined with MD simulations reported that lysozyme protein breaks the tetrahedral order of hydration water~\cite{russo2017pressure}. X-ray scattering experiment~\cite{bin2021wide} and MD simulations~\cite{melchionna2004water,dahanayake2018entropy} also detected reduced tetrahedral order in the protein hydration layer. Moreover, Merzel and Smith found that the hydration water of lysozyme is 15\% denser than bulk water~\cite{merzel2002first}. Since density is anticorrelated to the local structural order of water~\cite{shi2018impact}, the increased density supports the depletion of tetrahedral order in the protein hydration layer.

The effect of protein on the local structural ordering of hydration water has remained elusive so far~\cite{ball2017water}. The difficulty arises from the fact that neither the translational nor the rotational symmetry preserves at the protein/water interface, and thus, traditional structural descriptors targeting the tetrahedral order may not be suited for protein hydration water at the interface~\cite{accordino2012temperature}. In this work, we analyzed the water H-bond structure and applied a newly developed structural descriptor to protein hydration water. We find that the structural characterization focusing on the H-bond network unambiguously detects enhanced local structural ordering of the protein hydration water. This work not only opens a new door to the structural characterization of protein hydration water but also provides microscopic evidence supporting Kauzmann's seminal idea on the hydrophobic interaction.

\section{Methods}
In this study, we take the hen egg white lysozyme as the model protein, since it has been widely studied as an archetype protein in both experiments and simulations. The lysozyme protein contains 129 residues and the initial structure is obtained from the protein data bank (ID: 1IEE)~\cite{sauter2001structure}. The CHARMM36 force field~\cite{huang2013charmm36} was adopted to describe the interactions of protein and the water was modeled by the TIP4P/2005 model~\cite{abascal2005general}. A lysozyme protein was solvated in a cubic box of 63082 water molecules. Eight chloride ions were added to keep the charge neutrality of the system. The box is around $124 \times 124 \times 124$~\AA~with the periodic boundary condition applied in all directions. The system was equilibrated at 300~K and 1~bar for 5~ns and followed by another 2.1~ns NVT equilibration run at 300~K with the volume determined from the NPT run. Then a production run was performed in NVT ensemble at 300~K for 2.4~ns and the configurations were sampled every 0.2~ps. All the bonds with hydrogen atoms were constrained by the LINCS algorithm. A timestep of 2~fs was adopted for the simulations. The temperature and pressure were kept constant by using the Nose-Hoover thermostat and the Parrinello-Rahman barostat, respectively. The van der Waals and the electrostatic interactions in the real space were truncated at 12~\AA~and the electrostatic interactions in the reciprocal space were treated by the fast smooth particle-mesh Ewald method. Simulation of pure water was carried out in a system of 27000 TIP4P/2005 water molecules at 300~K for 2.4~ns with the other parameters the same as the protein simulations. All the simulations were performed by using the GROMACS (2019.4) package~\cite{Hess2008,abraham2015gromacs}.

\section{Results and discussion}

In pure water, molecules favor the tetrahedral arrangement of neighboring molecules. The degree of the tetrahedral order can be described by a parameter $q$ as~\cite{Chau1998,Errington2001}
\begin{equation}
	q=1-\frac{3}{8}\stackrel[i=1]{3}{\sum}\stackrel[j=i+1]{4}{\sum}\left(\cos\theta_{ij}+\frac{1}{3}\right)^{2},
	\label{eq:q}
\end{equation}
where $\theta_{ij}$ is the angle formed by two vectors connecting the central molecule and its nearest neighbors $i$ and $j$, and the summation runs over all the combinations of the four nearest neighbors. It takes a value of 0 and 1 for a random and a perfect tetrahedral configuration, respectively. The tetrahedral parameter $q$ has been widely used to characterize the tetrahedral order of protein hydration water~\cite{melchionna2004water,dahanayake2018entropy,russo2017pressure}. In pure water, the parameter $q$ is defined by using the oxygen atoms only. However, as pointed out by Accordino et al., the parameter $q$ may not be suited for water at the interface, because interfacial water may not often have four neighbors in the first coordination shell~\cite{accordino2012temperature}. The protein N and O atoms that are able to form H-bonds with water are often involved in the definition of $q$ to compensate for the loss of neighboring water molecules at the protein/water interface. The tetrahedral parameter $q$ targeting the rotational (tetrahedral) symmetry is determined solely by the angular distribution of neighbors. However, the presence of protein inevitably breaks the translational and rotational symmetry of the water arrangement at the interface.

There are many other structural descriptors focusing on the translational order of water, such as $d_5$~\cite{Cuthbertson2011} and local-structure index.~\cite{Shiratani1998}. These structural descriptors have been successfully applied to the characterization of the translational order of pure water. However, none of the above-mentioned structural descriptors, including the tetrahedral parameter $q$, consider the H-bond formation in their definitions. Since H-bond formation is the essential driving force for water structuring, characterization of the H-bond network in the protein hydration layer is crucial to reconcile the discrepancy in the structural description of protein hydration water.

\begin{figure}[t!]
	\begin{center}
		\includegraphics[width=8cm]{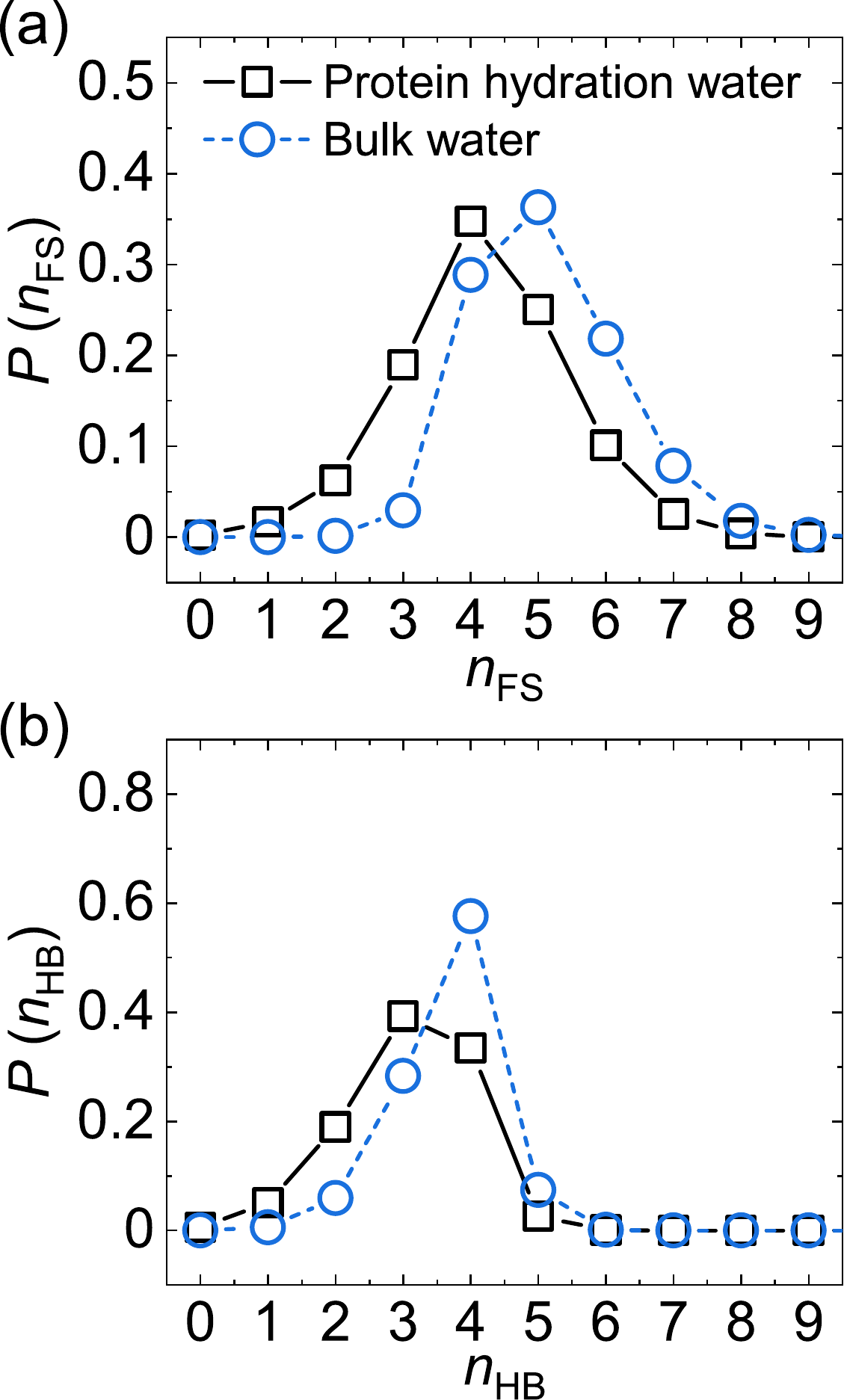}
	\end{center}
	\vspace{-6mm}
	\caption{The distribution of (a) the coordination number $n_\mathrm{FS}$ and (b) the number $n_\mathrm{HB}$ of H-bonded water neighbors per water molecule for protein hydration water (black squares) and bulk water (blue circles).}
	\vspace{-3mm}
	\label{fig:number}
\end{figure}

To find the protein hydration water, we adopted the cutoff method by which a water molecule is selected as protein hydration water if it is within 5~\AA~of at least one carbon atom of the protein~\cite{persson2018geometry}. Persson and coworkers have shown that this cutoff method is able to detect protein hydration water efficiently and accurately~\cite{persson2018geometry}. Figure~\ref{fig:number}(a) shows the distribution of coordination number $n_\mathrm{FS}$ (the number of water molecules in the first coordination shell) of protein hydration water and bulk water. We can see that the protein hydration water has a similar distribution shape compared to bulk water, but the peak position shifts from $n_\mathrm{FS}=5$ for bulk water to $n_\mathrm{FS}=4$ for protein hydration water. This shift of the distribution corresponds to the reduction of the water coordination number from 5.09 for bulk water to 4.18 for protein hydration water (Table~\ref{table:average}), which is ascribed to the confinement effect induced by the presence of protein. Figure~\ref{fig:number}(b) displays the distribution of number $n_\mathrm{HB}$ of H-bonded water neighbors per water molecule for protein hydration water and bulk water. Here, two water molecules are considered as H-bonded if their oxygen-oxygen distance is smaller than 3.5~\AA, and the H-O$\cdots$O angle is less than 30$^\circ$~\cite{Luzar1996hydrogen,Luzar1996effect}. The presence of protein not only shifts the peak position of the distribution from $n_\mathrm{HB}=4$ for bulk water to $n_\mathrm{HB}=3$ for protein hydration water but also changes the shape of the distribution. This can also be seen from Table~\ref{table:average} that each protein hydration water loses 0.58 H-bonds, compared to a loss of 0.91 neighbors on average, in the presence of protein, which suggests that the effect of protein is not only spatial confinement but also leads to the reorganization of water H-bond network.

\begin{figure}[t!]
	\begin{center}
		\includegraphics[width=8cm]{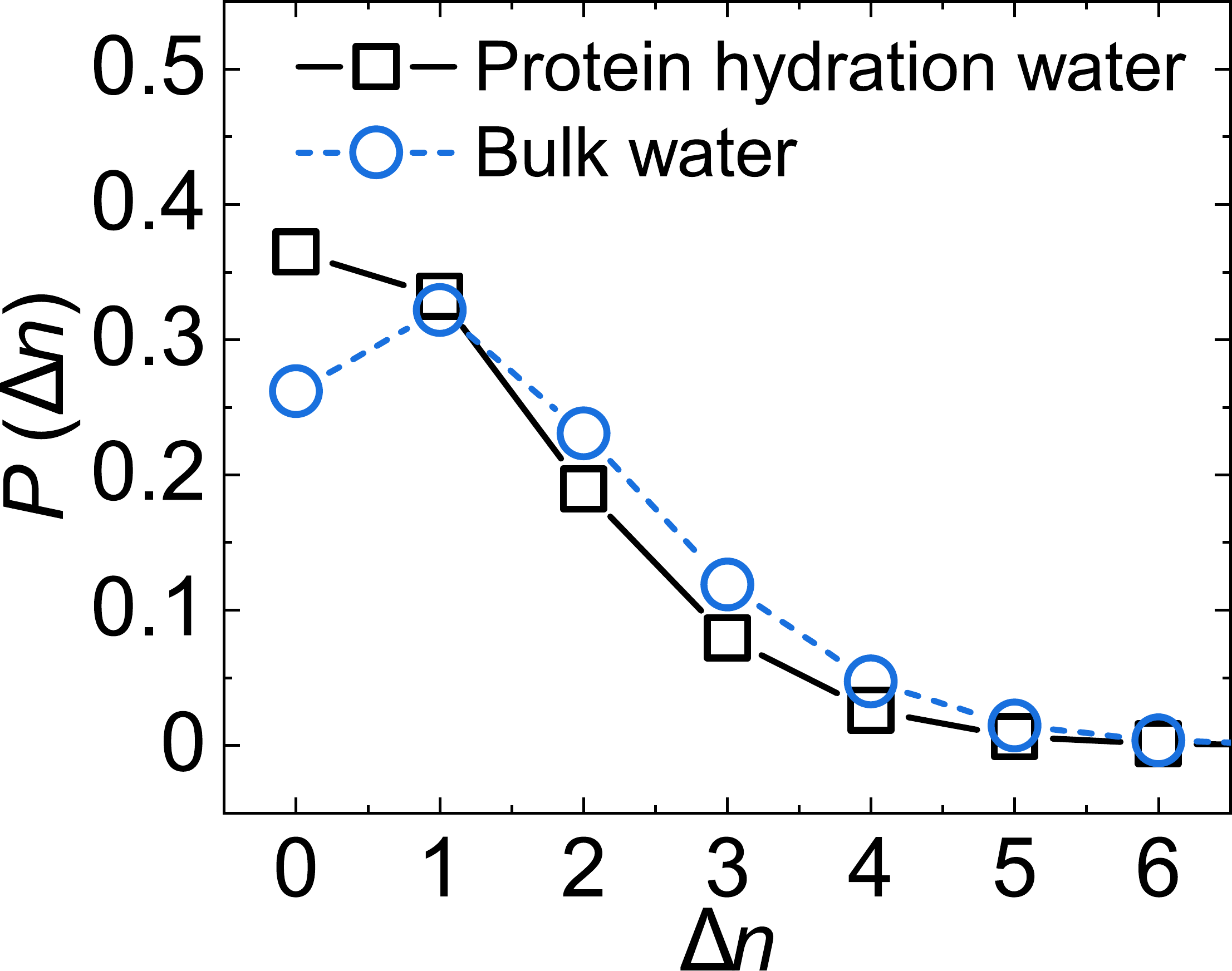}
	\end{center}
	\vspace{-6mm}
	\caption{The distribution of the number $\Delta n$ of non-H-bonded water neighbors in water's first coordination shell per water molecule for protein hydration water (black squares) and bulk water (blue circles).}
	\vspace{-3mm}
	\label{fig:dn}
\end{figure}

\begin{figure}[t!]
	\begin{center}
		\includegraphics[width=8cm]{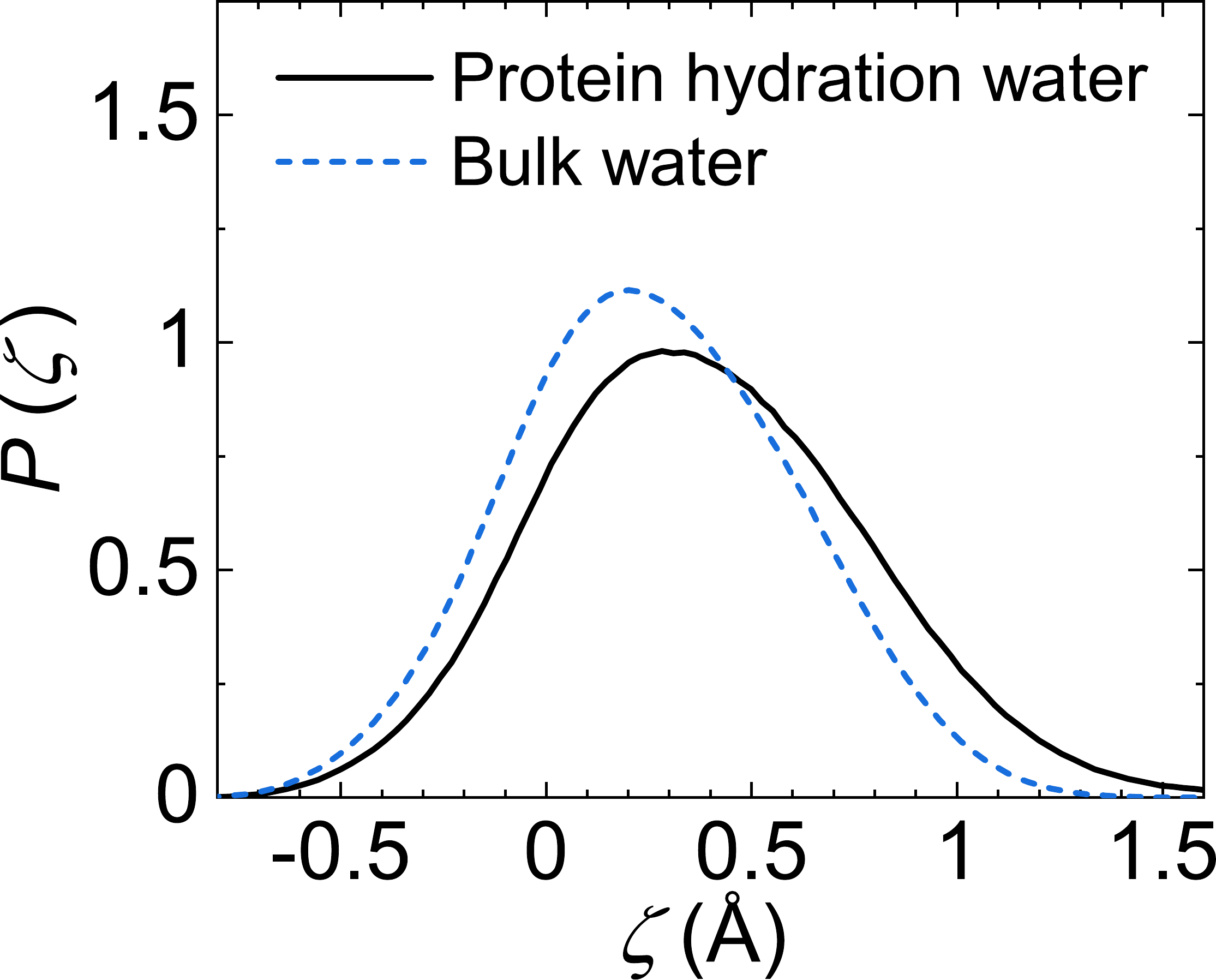}
	\end{center}
	\vspace{-6mm}
	\caption{The distribution of the structural descriptor $\zeta$ for protein hydration water (black solid line) and bulk water (blue dash line).}
	\vspace{-3mm}
	\label{fig:zeta}
\end{figure}

To characterize the effect of protein on water H-bond structure, we calculated the number of non-H-bonded water neighbors in water's first coordination shell which is defined by the following relation,
\begin{equation}
	\Delta n = n_\mathrm{FS} - n_\mathrm{HB}.
	\label{eq:dn}
\end{equation}
The distributions of $\Delta n$ for protein hydration water and bulk water are shown in Figure~\ref{fig:dn}. As we can see, protein significantly promotes the formation of a fully H-bonded first coordination shell ($\Delta n = 0$) by $\sim40\%$ and depletes the coordination shell with non-H-bonded neighbors, compared to bulk water. This result indicates that protein promotes the water H-bond structure in its hydration shell. Sciortino et al. have demonstrated that the presence of non-H-bonded molecules, which may be treated as "defects" in the first coordination shell, effectively enhances molecular mobility in liquid water~\cite{sciortino1991effect}. Thus, the promotion of the fully H-bonded first coordination shell should slow down the mobility of protein hydration water, which agrees with previous simulation and experimental results~\cite{fogarty2013,laage2017water}.

\begin{figure}[t!]
	\begin{center}
		\includegraphics[width=8cm]{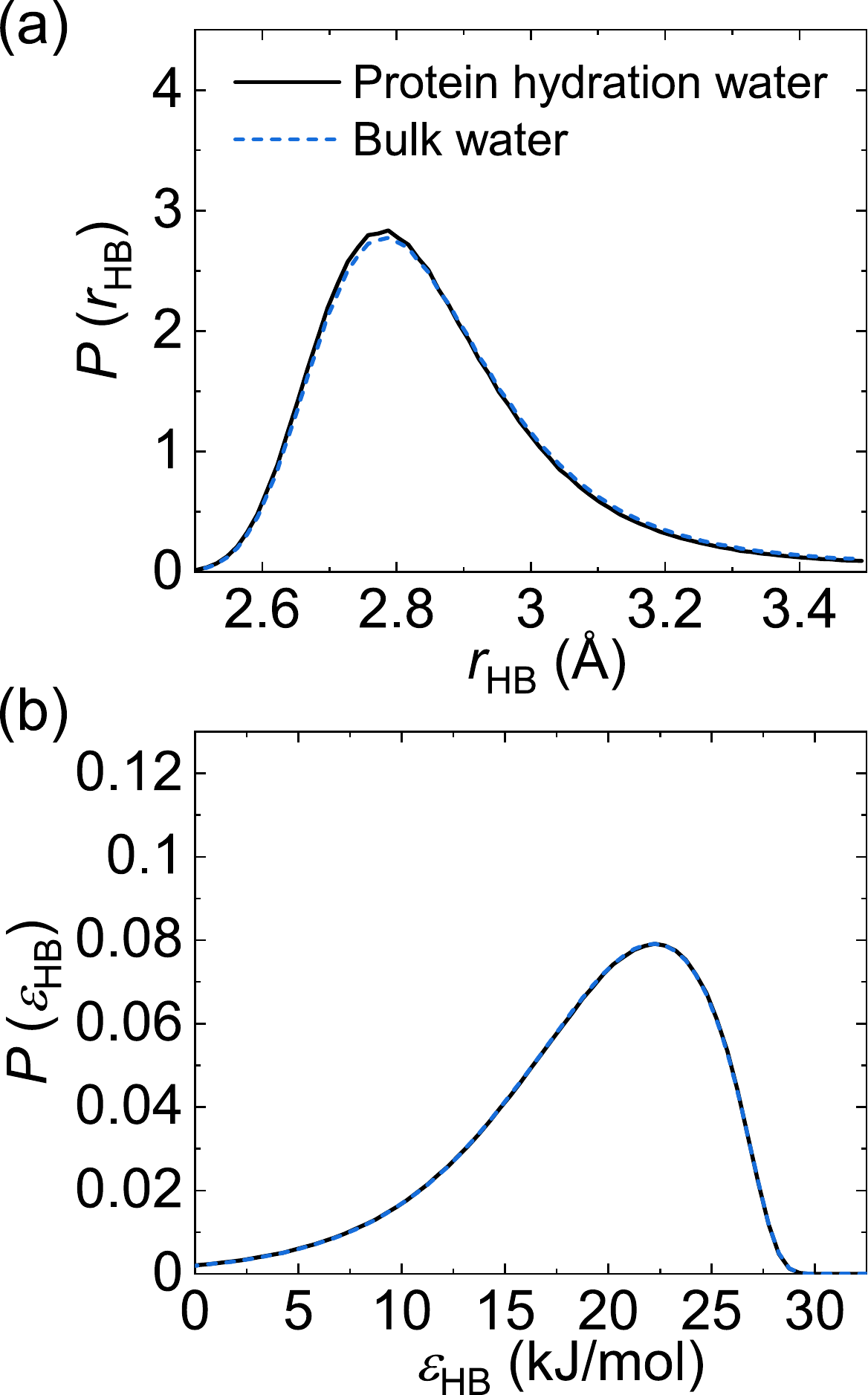}
	\end{center}
	\vspace{-6mm}
	\caption{The distribution of (a) the H-bond length $r_\mathrm{HB}$ and (b) the H-bond strength $\epsilon_\mathrm{HB}$ for protein hydration water (black solid line) and bulk water (blue dash line). In panel (b) the two distributions overlap.}
	\vspace{-3mm}
	\label{fig:hbond}
\end{figure}

\begin{table*}[t!]
	\caption{The average value of the structural descriptor $\zeta$, the coordination number $n_\mathrm{FS}$, the number of H-bonded water neighbors $n_\mathrm{HB}$, the number of non-H-bonded water neighbors $\Delta n$, the H-bond length $r_\mathrm{HB}$, and the H-bond strength $\epsilon_\mathrm{HB}$ of protein hydration water and bulk water obtained from our simulations. The standard deviations of the structural descriptors are shown in the parentheses.}
	\label{table:average}
	\vspace{3mm}
	\begin{tabular}{|c|c|c|c|c|c|c|}
		\hline 
		Water type & $\zeta$~(\AA) & $n_\mathrm{FS}$ & $n_\mathrm{HB}$ & $\Delta n$ & $r_\mathrm{HB}$~(\AA) & $\epsilon_\mathrm{HB}$~(kJ/mol) \\ 
		\hline
		Protein hydration water & 0.37~(0.40) & 4.18~(1.25) & 3.08~(0.94) & 1.10~(1.12) & 2.86~(0.17) & 18.91~(5.68) \\ 
		\hline
		Bulk water & 0.26~(0.35) & 5.09~(1.07) & 3.66~(0.73) & 1.43~(1.25) & 2.87~(0.18) & 18.92~(5.66) \\ 
		\hline
	\end{tabular} 
\end{table*}

\begin{figure*}[t!]
	\begin{center}
		\includegraphics[width=16cm]{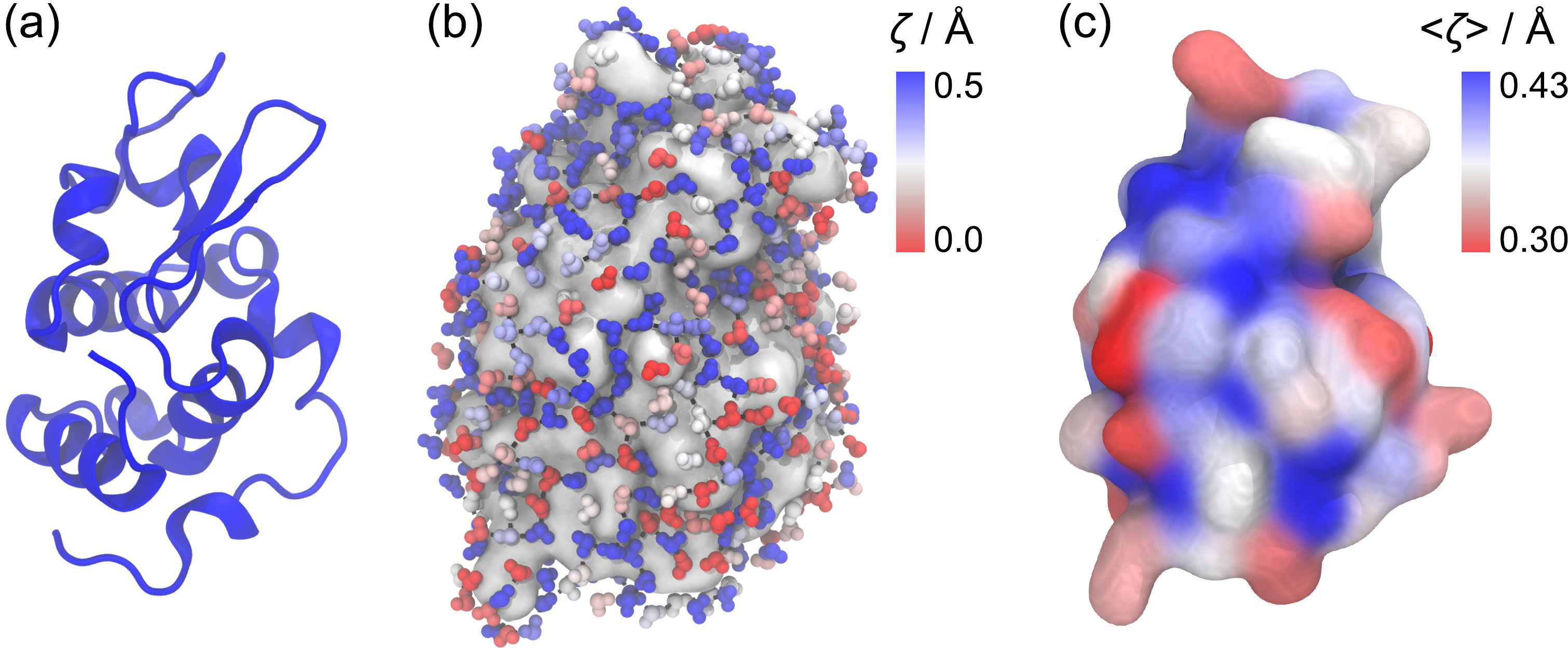}
	\end{center}
	\vspace{-6mm}
	\caption{(a) The snapshot of lysozyme protein in water at 300~K. (b) The snapshot of hydration water (small balls) on the surface of lysozyme protein (grey cloud) at 300~K. The water molecules are colored by the instantaneous value of the structural descriptor $\zeta$. (c) The surface of lysozyme protein at 300~K. The residues on the surface are colored by the value of $\langle \zeta \rangle$ that is defined by averaging the $\zeta$ value over all the water molecules in contact with the residue and over time. The blue and red color represents ordered and disordered water, respectively. The color bars for $\zeta$ and $\langle \zeta \rangle$ are shown in panel (b) and (c), respectively. The H-bonds formed between hydration water molecules are shown by black sticks.}
	\vspace{-3mm}
	\label{fig:snapshot}
\end{figure*}

Recently, Russo and Tanaka proposed a new structural descriptor $\zeta$ to characterize the local translational order of liquid water~\cite{russo2014understanding}. The descriptor $\zeta$ measuring the depth of non-H-bonded water penetrating into the first coordination shell is defined for each water molecule as
\begin{equation}
	\zeta = d_\mathrm{nhb} - d_\mathrm{hb},
	\label{eq:zeta}
\end{equation}
where $d_\mathrm{nhb}$ and $d_\mathrm{hb}$ are the distance from the closest non-H-bonded water and the distance from the furthest H-bonded water to the central molecule, respectively. A small $\zeta$ around 0 corresponds to a disordered structure with penetrated non-H-bonded molecule in the first coordination shell, whereas a relatively large $\zeta$ suggests a translationally ordered water structure with a fully H-bonded first coordination shell. Incorporating the H-bond information, the $\zeta$ parameter has been successfully exploited to characterize the local structural ordering in pure water~\cite{russo2014understanding,tanaka2019revealing,shi2018microscopic}.

Figure~\ref{fig:zeta} plots the distribution of $\zeta$ for protein hydration water and bulk water. Clearly, the protein hydration water has a broader distribution than bulk water, which may be attributed to the topological and chemical heterogeneities of the protein surface. Moreover, the $\zeta$ distribution for protein hydration water shifts towards large $\zeta$ value, compared to bulk water. Accordingly, the average $\zeta$ value increases by 42\% from 0.26~\AA~for bulk water to 0.37~\AA~for protein hydration water (Table~\ref{table:average}). The $\zeta$ distribution clearly demonstrates that the protein hydration water is structurally more ordered than bulk water, in agreement with the above analysis of the H-bond network (Figs.~\ref{fig:number} and~\ref{fig:dn}).

Besides H-bond structure, we also investigated the effect of protein on the length $r_\mathrm{HB}$ and strength $\epsilon_\mathrm{HB}$ of individual H-bond. Here, $r_\mathrm{HB}$ and $\epsilon_\mathrm{HB}$ are defined as the oxygen-oxygen distance and the interaction energy (in absolute value) of two H-bonded water molecules. Figure~\ref{fig:hbond}(a) and (b) show the distribution of $r_\mathrm{HB}$ and $\epsilon_\mathrm{HB}$, respectively. In contrast to the significant impact of protein on water H-bond structure, the presence of protein turns out to have negligible influence on either the length or the strength of water-water H-bond statistically. This result suggests that the protein promotes water's local structural ordering through the reorganization of the H-bond network, rather than perturbing the strength of individual H-bond.

The structure and dynamics of water in the near vicinity of the protein is rather heterogeneous~\cite{barnes2017spatially,heyden2019heterogeneity}. It has been shown that geometric topology~\cite{cheng1998surface,sheu2019surface}, charge distribution~\cite{cheng1999effect}, chemical nature~\cite{giovambattista2008hydrophobicity} and concentration~\cite{harada2012protein} of the protein all affect the structure of protein hydration water. In Figure~\ref{fig:snapshot} we show the snapshot of the lysozyme protein and its hydration layer. It can be seen that the cutoff method~\cite{persson2018geometry} accurately selected the hydration water at the protein/water interface. To illustrate the spatial heterogeneity of water's local structural order, we calculate the $\zeta$ parameter of each hydration water and show the instantaneous value of $\zeta$ by the color of water molecules in Figure~\ref{fig:snapshot} (b). As clearly indicated by $\zeta$ parameter, the local structural ordering of protein hydration water takes place heterogeneously on the protein surface. Moreover, the protein hydration water molecules with similar $\zeta$ values tend to aggregate into small patches on the protein surface, suggesting that the local structural ordering is not random but takes place in a cooperative manner. We note that the value of $\zeta$ fluctuates with time due to the thermal fluctuations of water structure at finite temperatures. Therefore, we calculated the average value of $\zeta$ for each residue, $\langle \zeta \rangle$, that is defined by averaging the $\zeta$ value over all the water molecules in contact with the residue and over time. Here, a water molecule is considered in contact with a residue if it is in the hydration shell of the protein and the residue is the closest one to that water molecule. The average value $\langle \zeta \rangle$ provides a measure of the degree of water structuring in the vicinity of each residue. We plot the spatial distribution of $\langle \zeta \rangle$ in Figure~\ref{fig:snapshot} (c). It can be seen that the structure of hydration water is indeed heterogeneous on the protein surface and this structural heterogeneity is strongly correlated with the residues on the protein surface. Understanding the origin of the spatial heterogeneity of water structuring on the protein surface and its link to the structure and chemical nature of the residues is of great interest for future study.

\section{Summary}
We have studied the effect of lysozyme protein on the structure of hydration water by all-atom MD simulations. Previous studies characterize the protein hydration water by structural descriptors focusing on either tetrahedral or translational order neglecting H-bond information. In this work, we have focused on the structure of water's H-bond network at the protein/water interface. We find that the protein facilitates the formation of a fully H-bonded first coordination shell of water in absence of any penetrating non-H-bonded molecules ("defects") on the protein surface. Moreover, the presence of protein tends to deplete the disordered water structure with non-H-bonded molecules in water's first coordination shell. Applying a newly developed translational structural descriptor $\zeta$ that explicitly takes H-bond formation into account, we find that the presence of protein promotes the $\zeta$ value by 42\% for the protein hydration water compared to bulk water. This result, together with the analysis of H-bond network, clearly demonstrates the significant development of the local structural order of water at the protein/water interface. This work highlights the essential role of H-bonding in the structural characterization of the interfacial water~\cite{shi2018microscopic} and provides clear microscopic evidence for the water structural ordering around the protein that underlies the essential hydrophobic interactions in biological systems~\cite{kauzmann1959some}. As an archetype protein, lysozyme contains various kinds of residues (polar, non-polar, positively charged, and negatively charged ones) and forms different types of protein structures ($\alpha$-helices, $\beta$-sheets, and loops). Thus, the results obtained from lysozyme protein are expected to be relevant in general for other proteins as well. This work provides new insights into the microscopic structural characterization of protein hydration water and is fundamental to the understanding of the solvation of biomolecules in water.

\vspace{1cm}
\noindent
{\bf Acknowledgements}
\noindent
We thank Prof. Jingyuan Li for fruitful discussions. This work was supported by the National Natural Science Foundation of China (Grant No. 12175196).

Correspondence and requests for materials should be addressed to R.S. (ruishi@zju.edu.cn).

\bibliography{lysWater}

\end{document}